\documentclass[12pt]{article}
\usepackage{fullpage}

\newcommand\be{\begin{equation}}
\newcommand\ee{\end{equation}}
\newcommand\bea{\begin{eqnarray}}
\newcommand\eea{\end{eqnarray}}
\newcommand\nn{\nonumber}

\newcommand\bdm{\begin{displaymath}}
\newcommand\edm{\end{displaymath}}

\def\pmb#1{\setbox0=\hbox{#1}%
  \kern-.025em\copy0\kern-\wd0
   \kern.05em\copy0\kern-\wd0
   \kern-.025em\raise.0433em\box0 }

\def\rd{{\rm d}}

\begin{document}

\title{\large {\bf  The origin of the algebra of quantum operators in the stochastic formulation of quantum mechanics\thanks{{\em Letters in Mathematical Physics} {\bf 3} (1979) 367--376. {\em Copyright}  \copyright \, 1979 {\em by D.\ Reidel Publishing Company, Dordrecht, Holland, and Boston, U.S.A.\/}}}}

\author{Mark Davidson\thanks{
Current Address: Spectel Research Corporation, 807 Rorke Way, Palo Alto, CA   94303
\newline  Email:  mdavid@spectelresearch.com, Web: www.spectelresearch.com}\\
\normalsize{\em}}

\date{}

\maketitle

\noindent {\bf Abstract}.   The origin of the algebra of the non-commuting operators of quantum mechanics is explained in the general F\'enyes-Nelson stochastic models in which the diffusion constant is a free parameter. This is achieved by continuing the diffusion constant to imaginary values, a continuation which destroys the physical interpretation, but does not affect experimental predictions.  This continuation leads to great mathematical simplification in the stochastic theory, and to an understanding of the entire mathematical formalism of quantum mechanics.  It is more than a formal construction because  the diffusion parameter is not an observable in these theories.

\section{Introduction}

The complementarity view of quantum mechanics dominates in most circles of modern physics.  The wave-particle dualism of this interpretation, and the assertion that a wave function describes the state of a particle completely has been a source of concern and mystery for many who have studied the subject.  The stochastic formulation offers a possible alternative to the complementarity view.  Quantum particles move along continuous trajectories in the stochastic picture, and quantum averages are ensemble averages over the space of trajectories.  The subject is in a state of development, and it is too early to tell whether the present stochastic models will lead to a satisfactory  explanation of quantum mechanical laws, but this prospect justifies further exploration.

A history of stochastic models of quantum mechanics has been given by Jammer \cite{jammer1}.  Fritz Bopp, whose numerous contributions are reviewed by Jammer \cite{jammer1}, has laid a philosophical and physical foundation for this theory.  Imre F\'enyes \cite{fenyes1} is credited with the first proof that Schr\"odinger's equation could be understood as a kind of diffusion equation for a Markov process.  Edward Nelson \cite{nelson1,nelson2} greatly elaborated on the work of F\'enyes, and put the theory on a much more rigorous footing.  At about the same time, Favella \cite{favella1} made strides in understanding the mathematics of quantum mechanics in terms of diffusion processes.  Since then, de la Pena-Auerbach \cite{auerbach1,auerbach2} has made numerous contributions.  An excellent review has been given by Diner and Claverie \cite{diner1}.  Related models have been proposed by Wiener and Siegel \cite{wiener1}, Della-Riccia and Wiener \cite{della1}, Bohm and Vigier \cite{bohm1}, Moyal \cite{moyal1}, and many others.

A continuum of F\'enyes-Nelson type models, each with different diffusion parameter, are possible \cite{davidson1}.  The diffusion parameter is constrained to be positive, but is otherwise undetermined.  This freedom allows one to chose the diffusion parameter at will.  In this paper, this freedom is exploited, and the diffusion parameter is continued to the complex plane.  Although the mathematics of the diffusion theory can be continued in this way, the physical interpretation of the theory is lost.  The mathematics simplify greatly at one special complex value of the diffusion parameter, and this is the reason for considering this continuation.

Within the  F\'enyes-Nelson framework, it is possible to construct a Hilbert space which plays essentially the same role as the Hilbert space of states in quantum mechanics.  It is also possible to introduce non-commuting operators for position, velocity, acceleration, etc.\ \cite{davidson2}.  For real values of the diffusion parameter, the commutation rules for these operators are different from the corresponding quantum  operators.  However, when the continuation to the complex plane is performed, one finds that the algebra of the familiar quantum operators may be recovered.  The various quantum expectations may be considered as analytic continuations to complex diffusion parameters of stochastic expectations.  The hypothesis that measurable quantities are independent of the diffusion parameter allows  this transformation, and makes it more than just  a formal construction.

\section{The algebra of quantum operators}

Consider a diffusion process defined by:
\be
\rd \vec x =\vec b (x,t) \rd t +\rd \vec W
\label{1}
\ee
where $\vec x$ is the position of a particle, and $\vec W$ is a Wiener process satisfying:
\be
E(\rd W_i (t) \rd W_j (t)) =2\nu \delta_{ij} \rd t,
\label{2}
\ee
where $\nu$ is a positive constant, the diffusion parameter, with dimensions (length)$^2$/time.  The process in eqn.\ (\ref{1}) was studied by Nelson \cite{nelson1,nelson2} who established a strong connection with quantum theory.

$\vec x$ is known to be a Markov process for sufficiently regular $\vec b$, but the most general regularity conditions are not known.  Therefore, complete mathematical rigor shall not be attempted here.  It will be assumed that $\vec b$ is sufficiently well behaved so that $\vec x$ is a Markov process.

Let $(\Omega,\Sigma, P)$ be the underlying probability space for $\vec x$, and for $\omega\in\Omega$, $x(t,\omega)$ is a sample trajectory as $t$ varies with $\omega$ fixed.  These sample trajectories are continuous almost surely if $\vec b$ is sufficiently regular, and this property shall be assumed.  The process $\vec x$ satisfies:
\be
\lim\limits_{h\to 0_+} E\left( \frac{\vec x(t+h) -\vec x(t)}{h} \biggl.\biggr| x(t) =x\right) =\vec b (x,t)
\label{3}
\ee
where the notation $E(\cdot |\cdot)$ denotes conditional expectation.  Also,
\be
\lim\limits_{h\to 0_+} \,\frac{1}{h} E((x_i (t+h) -x_i (t)) (x_j (t+h) -x_j (t)) | x(t) =x) =2\nu \delta_{ij}.
\label{4}
\ee
Since $\vec x$ is a Markov process, a Markov transition function may be defined:
\be
P(x, t; y,s) =\lim\limits_{d^3x\to 0} \,\frac{1}{d^3x}\, P(\vec x(t) \in d^3 x|x(s) =y), \quad t>s
\label{5}
\ee
which satisfies a Chapman-Kolmogorov  equation:
\be
P(x,t; y,s) =\int d^3 zP(x,t ; z, u) P(z,u; y,s), \quad t>u>s.
\label{6}
\ee
Continuity of the sample paths requires:
\be
\lim\limits_{t\downarrow s} P(x,t; y,s) =\delta^3 (x-y).
\label{7}
\ee
The backward equation of Kolmogorov may be derived, formally, by choosing $u$ in eqn.\ (\ref{6}) to be close to $s$.  Then one can expand:
\bea
P(x,t; z,u)&=&P(x,t; y,u) +(\vec z-\vec y) \cdot \vec\nabla_y P(x, t; y,u)+\nn\\
 &+& \frac{1}{2} (z-y)_i (z-y)_j \,\frac{\partial^2}{\partial y_i \partial y_j}\, P(x, t; y, u) +\ldots \, .
\label{8}
\eea
Substitution of eqn.\ (\ref{8}) into eqn.\ (\ref{6}), using eqns.\ (\ref{3}) and (\ref{4}) yields
\be
\left[ \frac{\partial}{\partial s} +\vec b (y) \cdot \vec\nabla_y +\nu\Delta y\right] \, P(x,t; y,s)=0
\label{9}
\ee
which is the backward equation. Differentiating eqn.\ (\ref{6}) with respect to $u$, using eqn.\ (\ref{9}), integrating by parts, and taking $u\to t$, one finds the forward equation:
\be
\left[ \frac{\partial}{\partial t} + \left( \vec\nabla_x \cdot \vec b\right) +\vec b \cdot \vec\nabla_x -\nu \Delta_x\right] \, P(x,t; y,s) =0.
\label{10}
\ee
Nelson \cite{nelson1,nelson2} also considers the time reversed process:
\bea
&&\rd\vec x= \vec b_* \rd t +\rd \vec W_*,\nn\\
&&\vec b_* =\lim\limits_{h\to 0_+} E\left( \frac{x(t) -x(t-h)}{h} \biggl.\biggr| x(t)=x\right),\nn\\
&& E(dW_{*i} \rd W_{*j})=2\nu\delta_{ij} \rd t.
\label{11}
\eea
He shows that, independent of any dynamical assumption, one has
\be
(\vec b -\vec b_* )/2 =\nu \vec \nabla \mbox{ ln } (\rho), \qquad \rho =\mbox{ probability density}.
\label{12}
\ee
He defines forward and backward time derivatives, $D$ and $D_*$, by
\bea
Df(x,t) &=&\lim\limits_{h\to 0_+} \,\frac{1}{h} \,E(f(x(t+h)) -f(x(t))|x(t) =x)\label{13}\\
D_*f(x,t) &=&\lim\limits_{h\to0_+} \,\frac{1}{h}\, E (f(x(t),t) -f(x(t-h), t-h)|
x(t) =x)
\label{14}
\eea
and $D$ and $D_*$ are found to be:
\be
D=\frac{\partial}{\partial t} +\vec b\cdot \vec \nabla +\nu \Delta, \qquad D_* =\frac{\partial}{\partial t} +\vec b_*\cdot \vec\nabla -\nu \Delta.
\label{15}
\ee
The mean acceleration is defined by
\be
\vec a =\frac{1}{2} [DD_* + D_*D]\vec x.
\label{16}
\ee
Equating $m\vec a$ to the force leads to  Schr\"odinger's equation
\be
m\vec a =-\vec\nabla V\to \left[-\frac{\hbar^2}{2m} \Delta +V\right] \exp (R+iS_N) =i\hbar \frac{\partial}{\partial t} \,\exp (R+iS_N),
\label{17}
\ee
where
\be
\vec\nabla \times \vec b =0, \, \nu =\frac{\hbar}{2m}, \quad \rho =\exp (2R), \quad \vec b=2\nu \vec \nabla (R+S_N).
\label{18}
\ee
The dynamical assumption, eqn.\ (\ref{17}), is not unique \cite{davidson1}.  If
\be
m\vec a +m(\beta/8) (D-D_*)^2\vec x =-\vec\nabla V,
\label{19}
\ee
then one obtains \cite{davidson1}:
\be
\left[-\frac{\hbar^2}{2m}\, \Delta +V\right] \exp (R+izS_N)=i\hbar \,\frac{\partial}{\partial t} \exp (R+izS_N),
\label{20}
\ee
provided that
\bea
&&\vec\nabla\times \vec b=0, \qquad \vec b =2\nu \vec \nabla (R+S_N), \qquad \rho =\exp (2R),\nn\\
&&\nu =z\frac{\hbar}{2m}, \quad z=1/\sqrt {1-\beta/2}.
\label{21}
\eea
In eqn.\ (\ref{20}),  $zS_N$ should be considered fixed as $z$ and $\nu$ range over their possible values.  Since $\beta$ is a free parameter, $\nu$ can take on any real value from 0 to $\infty$.  In order that $z$ be real, we must restrict $\beta<2$, from (\ref{21}).

Next, a Hilbert space is introduced, with operators corresponding to dynamical variables \cite{davidson2} (note that $\nu$ in Ref.\ 14 differs from $\nu$ here by a factor of 2).  Let ${\cal H}_t$ be the Hilbert space of complex functions, $f(x)$, $x\in R^3$, with inner product given by:
\be
(f, g) =E(f^*(x) g(x)) =\int d^3 x \rho (x,t) f^* (x) g(x).
\label{22}
\ee
Operators for $\vec{\dot x}$, $\vec{\ddot x}$, etc., shall now be defined.  First $\vec{\dot x}$:
\be
\vec{\dot x} f=\lim\limits_{\stackrel{\mbox{\scriptsize$u\uparrow t$}}{\mbox{\scriptsize$s\uparrow u$}}} \,\frac{\partial}{\partial u} \, E (\vec x (u) f(x(t)) |x (s) =x),\qquad t>u>s.
\label{23}
\ee
This definition is, for practical purposes, equivalent to the one used in Ref.\ 14.  The domain and range of $\vec{\dot x}$ will depend on the properties of the Markov process in question.  A detailed investigation of these will not be made.  Instead, regularity assumptions will be pointed out along the way to a formal derivation of $\vec{\dot x}$. $\vec{\dot x}$ may be evaluated, for a certain class of functions $f$, by using the Markov transition function:
\bea
&&\frac{\partial}{\partial u}\,E(\vec x(u)f(x(t))|x(s) =x) =\int \rd^3 y\rd^3 zf(z) \vec y\,\frac{\partial}{\partial u} \, P(y,u; x,s) \, P \, (z,t; y,u)
\nn\\
&&\quad =\int \rd^3 y\rd^3 xf (z) \vec y \left\{ P(z,t; y,u) \left[- \left(\vec \nabla_y \cdot \vec b (y,t) \right) -\vec b (y,t) \cdot \vec\nabla_y+\nu \Delta_y\right] P(y,u;x,s)+\right.\nn\\
&&\qquad + P(y,u; x,s) \left[ -\vec b (y,t) \cdot \vec \nabla_y -\nu \Delta_y\right] P(z,t: y,u)\left.\right\}
\label{24}
\eea
where it has been assumed that the order of integration and differentiation can be freely interchanged in (\ref{24}), and the forward and backward equations (eqns.\ (\ref{10}) and (\ref{9})) have been used to reexpress the $u$ derivative above.  Integration by parts yields:
\be
=\int\rd^3 y\rd^3 zf (z) P(z,t; y,u) \left[\left(\vec \nabla_y \cdot \vec b(y,t)\right) + \vec b (y,t) \cdot \vec \nabla_y -\nu \Delta_y \right] \, P(z,t ; y,u)
\label{25}
\ee

where $\left[\cdot,\cdot\right]$ denotes commutator. In the limit $u\uparrow t$, $P(z,t; y,u)\to \delta^3 (z-y)$, and therefore, assuming that this limit can be taken inside the integral:
\be
\vec{\dot x} f(x) =\lim\limits_{s\uparrow t} \int \rd^3 yf(y) \left[ \left( \vec\nabla_y \cdot \vec b (y,t)\right) +\vec b (y,t) \cdot \vec\nabla_y -\nu \Delta_y , y\right] P(y,t; x,s).
\label{26}
\ee

Another integration by parts yields:
\be
\vec{\dot x}  f(x)\lim\limits_{s\uparrow t} \int \rd^3 yP (y,t; x,s) \left[\vec y, -\vec b(y,t) \cdot \vec \nabla_y -\nu \Delta_y\right] f(y)=\left[\left(\vec b +\nu \vec\nabla\right) \cdot \vec\nabla, \vec x\right] f(x)
\label{27}
\ee
so that
\be
\vec{\dot x} =\vec b +2\nu \vec \nabla.
\label{28}
\ee
Operators for higher time derivatives are calculated in a similar fashion. One finds
\be
\vec x^{n+1} =\left[ \left(\vec b +\nu \vec\nabla\right) \cdot \vec \nabla, \vec x^n\right] + \frac{\partial}{\partial t}\, \vec x^n
\label{29}
\ee
where $\vec x^n$ denotes the $n$th time derivative:
\be
\vec x^n f(x) =\lim\limits_{\stackrel{\mbox{\scriptsize$u\uparrow t$}}{\mbox{\scriptsize$s\uparrow u$}}}\, \frac{\partial^n}{\partial u^n}\, E(\vec x(u) f(x(t)) | x(s) =x).
\label{30}
\ee
In particular, for $\vec{\ddot x}$, one finds
\bea
\vec{\ddot x} &=& \left[\left( \vec b +\nu \vec \nabla\right) \cdot \vec\nabla , \vec b +2\nu \vec \nabla\right] +\frac{\partial \vec b}{\partial t}\label{31}\\
\vec{\ddot x} &=&\frac{\partial}{\partial t} \, \vec b +\nu \left(\Delta \vec b\right) +\frac{1}{2} \left( \vec \nabla \vec b^2\right) +( \vec \nabla \times \vec b) \times (\vec b +2\nu \vec \nabla).
\label{32}
\eea
If $\vec\nabla \times \vec b =0$, as in eqn.\ (\ref{21}), then
\be
\vec{\ddot x} =\frac{\partial \vec b}{\partial t} +\nu \Delta \vec b+\frac{1}{2} (\vec \nabla \vec b^2).
\label{33}
\ee
It is desirable to express $\vec{\ddot x}$ in terms of the potential $V$.  To do this, using eqn.\ (\ref{21}), rewrite (\ref{33}) as:
\be
\vec{\ddot x} =\vec \nabla \left[ \exp (-R-S_N) \left(\biggl( 2\nu \frac{\partial}{\partial t} +2\nu^2 \Delta\biggr)  \exp \left(R+S_N\right)\right)\right].
\label{34}
\ee
Now, using eqns.\ (\ref{20}) and (\ref{21}), and after some tedious algebra, one finds:
\be
m\vec{\ddot x} =-\vec \nabla V+\vec \nabla \left(\frac{\hbar^2}{2m} +2m\nu^2\right) \frac{\Delta \sqrt\rho}{\sqrt \rho} .
\label{35}
\ee

The operators for $\vec x$ and $\vec{\dot x}$ do not commute.
\be
[\dot x_i, x_j] =2\nu \delta_{ij}; [x_i, x_j] =0; \quad [\dot x_i, \dot x_j]=2\nu (\partial_i b_j-\partial_j b_i),
\label{36}
\ee
where all velocities do commute if $\vec \nabla \times \vec b=0$, as in eqn.\ (\ref{21}).

If $F(x, \dot x)$ is any ordered polynomial function of the operators $\vec x$ and $\vec{\dot x}$, then one has
\bea
(g, F(x,\dot x)h) &=& \int\rd^3 x\rho (x,t) g^* (x) F(x, \vec b +2\nu \vec \nabla)h(x)\nn\\
&=&\int\rd^3 x \exp (R-S_N) g^*(x) F(x, 2\nu \vec \nabla)h(x)\exp (R+S_N).
\label{37}
\eea
This last expression shows that $\dot x$ takes a particularly simple for $(2\nu \vec \nabla)$ if the Hilbert space ${\cal H}_t$ is mapped onto a new Hilbert space, call it $I_t$, by the mapping $T:Tf=\exp (R+S_N)f$, and where the inner product on $I_t$, denoted by $((\cdot,\cdot))$ preserves the norm, i.e.:
\be
(f,g) =((Tf, Tg)).
\label{38}
\ee
Clearly, this requires
\be
((f,g)) =\int \rd^3 x \exp (-2S_N)f^* (x) g(x).
\label{39}
\ee
From (\ref{37}), it is clear that $T$ maps $\vec{\dot x}$ onto the operator:
\be
T\vec{\dot x} =2\nu \vec\nabla, \quad \mbox{ or }\quad T(\vec{\dot x} f) =2\nu \vec \nabla Tf.
\label{40}
\ee
The acceleration operator, $\vec{\ddot x}$, remains invariant under $T$ if $\vec\nabla \times \vec b=0$, because $\vec{\ddot x}$ is simply a multiplicative function on $x$ in this case (see eqn.\ (\ref{33})).  Higher time derivatives may be calculated by use of the formula:
\bea
T\vec x^{n+1} &=&\frac{\partial}{\partial t} \,T\vec x^n +\frac{1}{2m\nu} \left[ H, T\vec x^n\right];\nn\\
H&=& \frac{1}{2} \frac{(2m\nu)^2}{m} \Delta +V -\left(\frac{\hbar^2}{2m} +2m\nu^2\right) \frac{\Delta\sqrt \rho}{\sqrt \rho}
\label{41}
\eea
which follows by considering eqn.\ (\ref{29}) under the mapping $T$.  $T\vec x^n$ denotes the image of $\vec x^n$ under $T$, i.e.: $T(\vec x^n f)=(T\vec x^n)Tf$.  From here on, the following notation shall be used:
\be
\vec X^n =T\vec x^n
\label{42}
\ee
and it follows from the above that
\be
(g, F(x,\dot x)h) =((Tg, F(X, \dot X)Th)).
\label{43}
\ee
It is possible to define an operator $\vec X (t+s)$ by:
\be
\vec X (t+s) =\sum\limits^\infty_{h=0} \vec X^n (s^n/n!),
\label{44}
\ee
and it may be shown that
\be
E(x(t+s_1) \times \ldots \times x(t+s_n)) =((\exp (R+S_N) , X(t+s_1) \times \ldots \times X(t+s_n) \exp (R+S_N)),
\label{45}
\ee
where $s_i \leq s_{i+1}$.  Note that the operators of eqn.\ (\ref{44}) do not commute for different $s$, and therefore the ordering matters for the right-hand side of eqn.\ (\ref{45}).  Equation (\ref{45}) is essentially the Feynman-Kac formula.  Its derivation is tedious and will be omitted, but it can be proved by expressing the left-hand side in terms of the Markov transition function and using the forward and backward equations.

$H$ in eqn.\ (\ref{41}) depends on $\rho$, and this fact complicates the expression for $X(t+s)$.  Note that if $\nu$ is allowed to be imaginary, then
\be
\nu =\pm i\frac{\hbar}{2m} \to H=\frac{1}{2} \frac{\hbar^2}{m} \Delta +V
\label{46}
\ee
and the term involving $\rho$ in $H$ becomes zero.  This is a major simplification.  Although an imaginary value of $\nu$ is physically meaningless, we may allow it to be imaginary without affecting measurable results, since $\nu$ is a free parameter, and all measurable quantities are independent of $\nu$, or at least this is the hypothesis.

Consider eqn.\ (\ref{46}).  If $\nu$ is allowed to take on this imaginary value, then the commutation rules become:
\be
[\dot x_i, x_j] =\left[\dot X_i, X_j\right] =\pm i\hbar \delta_{ij}.
\label{47}
\ee
From eqn.\ (\ref{21}) one has in this case $z=\pm i$, and since $zS_N$ is fixed, $S_N$ becomes imaginary:
\be
zS_N =S; S_N =\mp iS; \quad S \mbox{ real}.
\label{48}
\ee
The fact that $S_N$ is imaginary affects the inner product on the space $I_t$.  Equation (\ref{39}) is no longer applicable, but rather one finds:
\be
((f, g)) =\int \rd^3 xf^* g
\label{49}
\ee
so that in this case, $I_t$ is just $L^2$.  The operator $\vec{\dot X}$ becomes:
\be
\vec{\dot X} =\pm i\hbar /m\vec\nabla; \quad \vec P =m\vec{\dot X} =\pm i\hbar \vec\nabla
\label{50}
\ee
and this agrees with the familiar quantum mechanical result if the minus sign is chosen above.  The equation for time differentiation, eqn.\ (\ref{42}), becomes:
\be
\vec X^{n+1} =\frac{\partial}{\partial t} \vec X^n \mp \frac{i}{h} \left[H, \vec X^n\right] ; \quad H=-\frac{\hbar^2}{2m} \Delta +V
\label{51}
\ee
which agrees with the quantum mechanical result for the analogous operators.

The operators $X(t+s)$ are found to be the familiar Heisenberg position operators:
\be
X(t+s) =\exp \left( \mp \frac{i}{\hbar} Hs\right) X\exp \left( \pm \frac{i}{\hbar}Hs\right),
\label{52}
\ee
where it has been assume that $V$ does not depend explicitly on time in arriving at (\ref{52}).  The Feynman-Kac formula (eqn.\ (\ref{45})) becomes:
\bea
E_c (x(t+s_1) \times \ldots \times x(t+s_n)) &=& \int \rd^3 x \exp (R\pm iS)\times \ldots\nn\\
&&\ldots \times X(t+s_1) \times \ldots \times X(t+s_n)\exp (R\mp S_N),
\label{53}
\eea
where $s_i\leq s_{i+1}$, and where $E_c$ denotes a continuation, to complex $\nu =\pm i\hbar /2m$ (either sign is possible), of the  Markov expectation.  $E_c$ is in general complex, but not directly measurable.  If all of the times in (\ref{53}) are the same time, then $E_c$ becomes real, and equal to the real Markov expectation, which in this case is independent of $\nu$.

It is clear that the mathematical formalism of ordinary quantum mechanics is recovered if the minus sign of eqn.\ (\ref{46}) is chosen.  The interpretation is similar also.  Care must be taken to compare the theory with measurable quantities, thereby avoiding the ambiguities of an interpretation of complex expectations in eqn.\ (\ref{53}).  The underlying reason for the algebra of quantum operators is explained, however, in the stochastic formulation of quantum mechanics.  It has been derived here from the postulate that quantum mechanics is equivalent to a class of Markov processes with diffusion constant a free parameter.  The continuation to imaginary $\nu$ which leads to complex expectations is simply a convenient artifice, which facilitates calculations without affecting measurable results.

\section{Conclusion}

In the stochastic formulation of quantum mechanics, simplicity is achieved by exploiting the indeterminate nature of the diffusion constant, and choosing it to be a particular imaginary value.  Although the physical interpretation is lost, or at least obscured, by this, experimentally measurable averages are not affected, or at least this is the postulate which justifies the continuation, and it does not seem obviously false.  Perhaps a way to measure $\nu$ will someday be found, but this would require a measurement which goes beyond the ordinary predictions of quantum theory.

The central question, not discussed here, is that of the origin of the diffusion laws underlying quantum theory.  Several models have been proposed \cite{auerbach3}--\cite{davidson3}.  Although the definitive explanation has not yet been found, one possibility seems to stand out:  quantum mechanics may arise out of the interaction of charged particles with random forces in the vacuum, and with radiative forces playing an important role.  This random force may be due, at least in part, to the existence of the random radiation of stochastic electrodynamics \cite{boyer1}.

\bigskip

\noindent {(\em  Received June 1, 1979)}

\end{document}